\documentclass[aps,prb,twocolumn,showpacs,floatfix]{revtex4}

\usepackage{graphicx}
\graphicspath{{figs/}}
\begin{document}
\title{A theoretical study of thermal conductivity in single-walled boron nitride nanotubes}
\author{Jin-Wu~Jiang}
    \altaffiliation{Electronic address: phyjj@nus.edu.sg}
    \affiliation{Department of Physics and Centre for Computational Science and Engineering,
             National University of Singapore, Singapore 117542, Republic of Singapore }
\author{Jian-Sheng~Wang}
    \affiliation{Department of Physics and Centre for Computational Science and Engineering,
                 National University of Singapore, Singapore 117542, Republic of Singapore }

%\date{22 December 2009}
\date{\today}
\begin{abstract}
We perform a theoretical investigation on the thermal conductivity of single-walled boron nitride nanotubes (SWBNT) using the kinetic theory. By fitting to the phonon spectrum of boron nitride sheet, we develop an efficient and stable Tersoff-derived interatomic potential which is suitable for the study of heat transport in sp2 structures. We work out the selection rules for the three-phonon process with the help of the helical quantum numbers $(\kappa, n)$ attributed to the symmetry group (line group) of the SWBNT. Our calculation shows that the thermal conductivity $\kappa_{\rm ph}$ diverges with length as $\kappa_{\rm ph}\propto L^{\beta}$ with exponentially decaying $\beta(T)\propto e^{-T/T_{c}}$, which results from the competition between boundary scattering and three-phonon scattering for flexure modes. We find that the two flexure modes of the SWBNT make dominant contribution to the thermal conductivity, because their zero frequency locates at $\kappa=\pm\alpha$ where $\alpha$ is the rotational angle of the screw symmetry in SWBNT.
\end{abstract}

\maketitle

\pagebreak

\section{introduction}
Recently, experimentalists have accomplished significant progress with the synthesis of boron nitride (BN) nano materials, including the hybridized boron nitride and graphene,\cite{CiL} the boron nitride thin film,\cite{ShiY} and the boron nitride nanoribbon.\cite{ZengH} Prior to these breakthroughs, the BN nano materials have drawn continuous attention over past decades. The optical transitions in the single-walled boron nitride nanotubes (SWBNT) were measured by the optical absorption spectroscopy\cite{Lauret} or the spatially resolved electron energy loss spectroscopy.\cite{Arenal} The giant Stark effect in SWBNT was observed by the bias dependent scanning tunneling microscopy and scanning tunneling spectroscopy.\cite{Ishigami} The lattice dynamics properties have been investigated by first-principles calculation\cite{Kern,Wirtz,Kunstmann,Hamdi} or the valence force field model.\cite{Popov,Jeon,Michel} For the thermal transport in SWBNT, both experimental\cite{ChangCW2005,ChangCW2006} and theoretical\cite{Savic,Stewart} results have confirmed the importance of isotopic doping. The theoretical study of the thermal transport in SWBNT is mainly in the ballistic region where Landauer formula is applied with phonon dispersion obtained from lattice dynamics calculations. In these studies, the phonon-phonon scattering is ignored in the ballistic region, while the effect of phonon-phonon scattering on the thermal conductivity in carbon-based nano materials has been investigated by the Boltzmann equation approach.\cite{CaoJX,GuY,Nika,Lindsay2010} However, it is still a blank field for the the thermal conductivity of SWBNT. One of the main objectives of present paper is to investigate the effect of the phonon-phonon scattering on the thermal conductivity of SWBNT.

In this paper, we study the effect of the boundary and three-phonon scattering on the thermal conductivity in SWBNT. In our investigation, we calculate all required physical quantities from the Tersoff plus universal force field out-of-plane (Tersoff+UFFOOP) potential. The parameters of this potential are fitted to the phonon spectrum in sp2 BN sheet. It preserves the efficiency and stability of the original Tersoff potential. We find the selection rules for three-phonon process by using the helical quantum numbers $(\kappa,n)$ which are assigned to the screw and rotation symmetries in the line group of the SWBNT. The helical quantum numbers are conserved in the three-phonon scattering process and can distinguish the flexure modes in SWBNT from other acoustic modes. We find that the thermal conductivity of SWBNT shows diverging behavior with length as $\kappa_{\rm ph}\propto L^{\beta}$, where the exponent $\beta(T)$ decreases exponentially with the increase of temperature $T$. The thermal conductivity is dominated by the flexure modes, which can be clearly interpreted with the help of the helical quantum numbers.

The present paper is organized as follows. In Sec.~II, we find the selection rules for the three-phonon scattering process. Sec.~III is devoted to the interaction potential. Sec.~IV shows formulas for the phonon life time and thermal conductivity. In Sec.~V, calculation results are presented and discussed. The paper ends with a summary in Sec.~VI.

\section{symmetry selection rules}
In the scattering process, symmetry selection rules impose strong constraints on the symmetry properties of particles which participate in the process. Resulting from the scalar property of the scattering operator in the three-phonon process,\cite{Ziman} the selection rules reflect directly the conservation of quantum numbers corresponding to all symmetry operations in the system. First of all, the energy conservation leads to:
\begin{eqnarray}
\omega + \omega' = \omega''.
\end{eqnarray}
We will only exhibit selection rules for combining process. They can be obtained analogously for splitting process and will be listed at the end of this section. Besides the energy conservation, the other selection rules are governed by the symmetry group of the system. For chiral, armchair, and zigzag SWBNT($n_{1}$,$n_{2}$), the symmetry groups are the first, fourth, and eighth class of line group.\cite{Damnjanovic,Alon} The chiral SWBNT have the screw symmetries and the pure rotational symmetries with generators $S(\alpha,h)$ and $C_{N}^{1}$. The SWBNT are rotated around $z$ axis for $\alpha$ and translated by $h$ in the $z$ direction after the operation of $S(\alpha,h)$. $z$ is set to be the rotational axis of SWBNT. $\alpha$ and $h$ are related to $n_{1}$ and $n_{2}$.\cite{White1993} The rotational angle for $C_{N}^{1}$ is $2\pi/N$, where $N={\rm gcd}(n_{1},n_{2})$ is the greatest common divisor of $n_{1}$ and $n_{2}$. The armchair or zigzag SWBNT also have the screw and pure rotational operations. Besides, there are additional reflection symmetries $\sigma_{h}/\sigma_{v}$ in armchair/zigzag SWBNT. In the irreducible representations of the line group, the quantum numbers corresponding to the screw and rotation operations are a set of helical quantum numbers $(\kappa, n)$, with $\kappa\in(-\pi/h, \pi/h]$ and $n=(-N/2, N/2]$.\cite{Dobardzic} Using the group theory, one finds that the selection rules in chiral SWBNT for three-phonon process is merely to guarantee the conservation of helical quantum numbers $(\kappa, n)$. In armchair or zigzag SWBNT, the symmetry property is different for phonons in the center or at the edge of the Brillouin zone (BZ), i.e $(\kappa,n)=$ $(0,0)$, $(\pm \pi/h,0)$, or $(0,\pm N/2)$ for even $N$. Those phonons carry the information of the reflection symmetries; thus, they are of higher symmetry than the other phonons. From group theory, higher symmetry will lead to more complicate selection rules. However, those phonons of higher symmetry are unable to carry heat energy in the thermal transport, because they either correspond to a rigid movement of the system or have zero phonon velocity. It should be noted that the other long-wave acoustic modes make important contribution, except the mode with $(\kappa,n)=(0,0)$. After ignoring those phonons, the selection rules in armchair or zigzag SWBNT turn to be the same as that of the chiral SWBNT. Consequently, the selection rules in all SWBNT are:
\begin{eqnarray}
\kappa + \kappa' &=& \kappa'' \bmod \frac{2\pi}{h},\nonumber\\
n + n' &=& n'' \bmod N.
\end{eqnarray}
We emphasize that the physical consequence of these selection rules is to conserve the quantum numbers $(\kappa, n)$ corresponding to the screw and pure rotational symmetries of the SWBNT. Besides the simplicity in their selection rules, $(\kappa, n)$ have several advantages. To the best of our knowledge, present paper is the first work to apply the helical quantum numbers which correspond to the actual symmetries of the SWBNT in the study of phonon scattering process in nanotubes. All existing literatures use the linear quantum numbers $(k,m)$,\cite{SaitoR} attributed to the pure translational and rotational operations with generators $T_{\tilde{q}h}$ and $C_{q}^{1}$. $q$ is the number of atom pairs in a big translational unit cell and $\tilde{q}=q/N$. The operation $T_{\tilde{q}h}$ translates the SWBNT by $\tilde{q}h$ in $z$ direction.
\begin{figure}[htpb]
  \begin{center}
    \scalebox{1.0}[1.0]{\includegraphics[width=8cm]{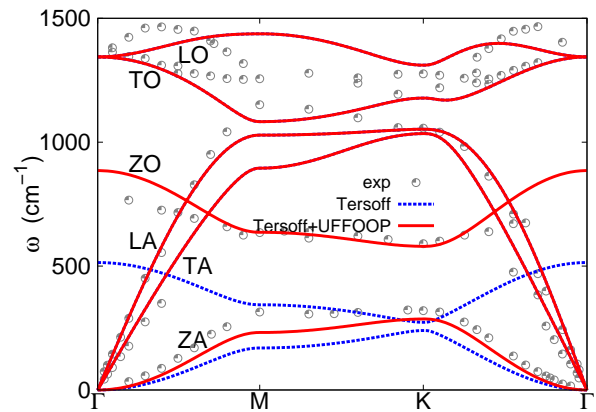}}
  \end{center}
  \caption{(Color online) Phonon spectrum in BN sheet calculated from Tersoff potential, Tersoff+UFFOOP potential are compared with experimental results from Ref.~\onlinecite{SerranoJ}. The UFFOOP potential enhance merely the two out-of-plane (ZA and ZO) vibration modes.}
  \label{fig_bn_sheet}
\end{figure}
 All of those pure translational and rotational operations together form a subgroup of the line group. Although $T_{\tilde{q}h}$ is a symmetry operation in the SWBNT, the $C_{q}^{1}$ is not a symmetry operation of SWBNT. As a result, the linear quantum numbers $(k,m)$ are not conserved in the three-phonon process. The relations between linear and helical quantum numbers are $n=m~{\rm mod}~N$, and $\kappa =  k + m\frac{\alpha}{h}+j\frac{2\pi}{h}$ where $j$ is an integer to keep $\kappa$ in its BZ. These relations lead to an additional constraint on the selection rules of $(k,m)$, which has been originally pointed out by Dobardzic {\it et~al.}\cite{Dobardzic}, and demonstrated by Lindsay {\it et~al.}\cite{Lindsay2010}

Analogously, we can obtain the following selection rules for splitting process:
\begin{eqnarray}
\omega &=& \omega' + \omega'',\nonumber\\
\kappa &=& \kappa' + \kappa'' \bmod \frac{\pi}{h},\\\nonumber
n &=& n' + n'' \bmod N.
\end{eqnarray}

\section{interatomic potential}
The interaction for the sp2 BN sheet or SWBNT can be described by valence force field models in the linear approximation.\cite{Popov,Jeon,OuyangT} This linear model can be adopted to study the ballistic phonon transport without phonon-phonon scattering. To investigate the thermal transport beyond the ballistic region, it is necessary to apply a more realistic interatomic potential which can include the nonlinear effect and bond reaction phenomenon. Several sets of parametric Tersoff bond order potentials have been applied in the molecule dynamics simulation of various properties for BN materials.\cite{Sekkal,Matsunaga,Albe} However, these parameters are not fitted to the sp2 BN sheet structure. Especially, they are not good in the description of the phonon spectrum of BN sheet, which is crucial for the investigation of thermal transport in present work. It is important to fit a set of Tersoff parameters to the phonon dispersion of sp2 BN sheet, as there is no experimental values for nonlinear properties of BN sheet. The fitting procedure is realized in following steps. Firstly, the phonon dispersion in BN sheet is quite similar to that of the graphene, because these two materials have similar honeycomb structure. So we start with the Tersoff parameters for carbon,\cite{Tersoff} with parameters $A$ and $B$ in the Tersoff potential rescaled by a factor of 0.3. Fig.~\ref{fig_bn_sheet} (dash line, blue online) shows that this set of Tersoff parameters give a good description of the four in-plane vibrations: longitudinal acoustic (LA), transverse acoustic (TA), longitudinal optical (LO), and transverse optical (TO) modes. However, it results in much lower frequencies for the two out-of-plane vibrations: $z$ acoustic (ZA) and $z$ optical (ZO) modes. We find that it is impossible to account for the in-plane and out-of-plane vibrations simultaneously through adjusting the parameters in Tersoff potential where the bond order function depends on the bond length and in-plane bond angles. The in-plane transverse vibration is captured by the in-plane bond angle term. The in-plane longitudinal vibration and out-of-plane vibration are coupling together in the bond length term. As a result, it is impossible to modify Tersoff parameters to independently affect the out-of-plane vibration while keep in-plane longitudinal vibration unchanged. Brenner {\it et~al.} solved this problem by generalizing the bond order function to be dihedral angle dependent,\cite{Brenner} which can treat the out-of-plane vibrations separately. However, this is a numerically expensive generalization. Particularly for large systems, the numerical simulation using Brenner potential becomes much slower than the Tersoff potential. Actually, this problem can be settled down by a more efficient and straightforward approach. We introduce a concise potential to manipulate the out-of-plane vibration separately. It should be noted that it is an attractive way to describe each degree of freedom by a separate term, as this can provide clear physical picture for each phonon branch. Among various potentials, we find that the most stable one is the UFFOOP potential\cite{Gale}: $V=C_{0}(C_{1} + C_{2}\cos(\phi) + C_{3}\cos(2\phi))$, where $\phi$ is the dihedral angle. The four optimized parameters $C_{i}$ are shown in Table~\ref{tab_uffoop}. Fig.~\ref{fig_bn_sheet} (solid line, red online) shows that UFFOOP only enhances the out-of-plane vibration and does not affect the in-plane vibration at all in the planar BN sheet. For tubes, especially those with small diameters, the dihedral angle $\phi$ depends on the curvature of the tube, so the out-of-plane and in-plane vibrations are mixing together by the UFFOOP potential. The phonon spectrum calculated from Tersoff+UFFOOP potential agrees well with the experimental values. It should be noted however that the theoretical frequencies for LO and TO modes do not coincide with experimental data in detail, since these two modes are more sensitive to long-range interactions compared with acoustic branches.\cite{Michel} The long-range interactions originate from the charge polarization of boron and nitride atoms in this ionic system. Especially, the crossing of TO and LO branches can not be repeated by the Tersoff+UFFOOP potential. However, we can safely ignore the effect of long-range interaction in the study of phonon thermal transport, because the LO and TO modes only contribute to thermal conductivity through providing new channels for phonon-phonon scattering of other phonons due to their low phonon velocity. At this point, we would like to point out two distinct features of the Tersoff+UFFOOP potential. Firstly, it is of high efficiency and stability which inherits from the original Tersoff potential for carbon system. We have tested that the SWBNT are still stable in molecular dynamics simulation at 2000 K after $10^{10}$ steps. Secondly, this potential can give a good description simultaneously for the in-plane and out-of-plane vibrations of the sp2 BN system; thus suitable for the study of thermal transport in these structures.
\begin{table}[t]
\caption{Parameters of the four-body UFFOOP potential: $V=C_{0}(C_{1} + C_{2}\cos(\phi) + C_{3}\cos(2\phi))$. The second/third lines are for configuration with B/N in the center of other three N/B atoms. $C_{0}$ is in eV. Other three parameters are dimensionless.}
\label{tab_uffoop}
\begin{ruledtabular}
\begin{tabular}{ccccc}
bond type & $C_{0}$ & $C_{1}$ & $C_{2}$ & $C_{3}$\\
\hline
BNNN & -0.776387 & 1.000000 &  1.473839 & 0.424507\\
NBBB & 0.236178  & 1.000000 & -0.373869 & -0.223418\\
\end{tabular}
\end{ruledtabular}
\end{table}

\section{phonon life time and thermal conductivity}
The thermal conductivity can be calculated by following formula:\cite{GuY}
\begin{eqnarray}
\kappa_{\rm ph} = \frac{1}{V} \sum_{\kappa,n,\sigma} \tau_{\kappa,n}^{\sigma}C_{\rm ph}(\omega)v_{\kappa,n,\sigma,z}^{2},
\label{eq_conductivity}
\end{eqnarray}
where $V$ is the volume of the system. The thickness of the SWBNT is considered to be the inter-layer spacing (3.35~{\AA}) in the hexagonal BN multi-layers.\cite{ZengH} $\sigma$ is the polarization index of phonon dispersion. The phonon heat capacity is given by $C_{\rm ph}=k_{B}x^{2}e^{x}/(e^{x}-1)^{2}$ with $x=\hbar\omega/(k_{B}T)$. $v_{\kappa,n,z}^{\sigma}$ is phonon velocity in the direction of thermal current. Under the single mode relaxation time approximation, the relaxation rate for phonon mode $(\kappa, n, \sigma)$ due to three-phonon scattering process is:\cite{Khituna}
\begin{eqnarray}
\frac{1}{\tau_{\rm ps}} & = & \sum_{\kappa',n',\sigma'}2|C_{3}|^{2}\frac{\hbar}{M^{3}\omega\omega'\omega''}\pi\delta\left(\Delta\omega\right)N(\omega',\omega''),
\end{eqnarray}
where the coefficient $|C_{3}|^{2}  =  \frac{4\gamma^{2}}{3A}\times\frac{M^{2}}{v_{g}^{2}}\times\omega^{2}\omega'^{2}\omega''^{2}$. To simplify the notation, we have introduced a single prime to denote that the quantity corresponds to mode $(\kappa',n',\sigma')$. Similarly, a double prime means that the quantity corresponds to the mode $(\kappa'',n'',\sigma'')$.
\begin{figure}[htpb]
  \begin{center}
    \scalebox{1.0}[1.0]{\includegraphics[width=8cm]{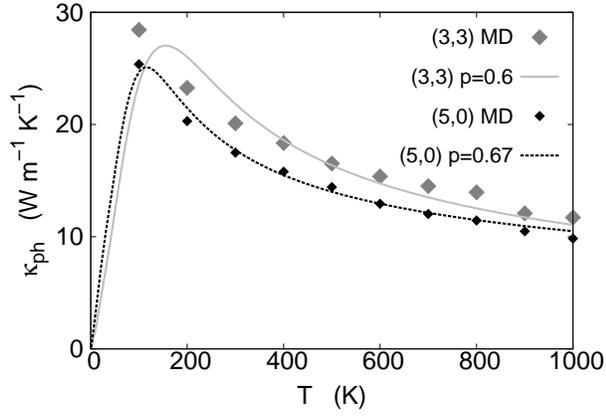}}
  \end{center}
  \caption{Fit specularity parameter $p$ to the thermal conductivity from molecular dynamics simulation.}
  \label{fig_pfitmd}
\end{figure}
 The two modes $(\kappa',n',\sigma')$ and $(\kappa'',n'',\sigma'')$ are related to each other through the selection rules in Sec.(II). $A$ is the total number of atoms in the system. $M$ is the atomic mass of a single atom. For the combining process, $\Delta\omega=\omega+\omega'-\omega''$ and $N(\omega',\omega'')=N_{0}'-N_{0}''$. The distribution function is $N_{0}=1/(e^{x}-1)$. For the splitting process, $\Delta\omega=\omega-\omega'-\omega''$ and $N(\omega',\omega'')=N_{0}'+N_{0}''+1$. $\gamma$ is the Gr\"uneisen parameter of mode $(\kappa,n,\sigma)$. It is important to use mode-specific Gruneisen parameters. This has been confirmed by Nika {\it et~al.} in the calculation of thermal conductivity of 2D graphene, which shows that the result from mode-dependent Gr\"uneisen parameters is in better agreement with experiment.\cite{Nika} $\gamma$ is calculated by $\gamma=-(V/\omega)(\partial\omega/\partial V)$ in the 3D structure, and by $\gamma=-(S/\omega)(\partial\omega/\partial S)$ in 2D system with $S$ as the area. In the quasi-one-dimensional SWBNT system, we calculate it by $\gamma=-(L/\omega)(\partial\omega/\partial L)$. The summation over $\kappa'$ in the relaxation rate can be changed into an integral, as the SWBNT are usually very long. The relaxation rate is then obtained by following formula:\cite{GuY}
\begin{eqnarray}
\frac{1}{\tau_{\rm ps}} & = & \left(\frac{4}{3\rho_{L}}\right)\left(\frac{\hbar\omega\gamma^{2}}{v_{z}^{2}}\right)\sum_{n'\sigma'}^{\prime}\frac{1}{v_{g}}\omega'\omega''N\left(\omega',\omega''\right),
\label{eq_pslifetime}
\end{eqnarray}
where $\rho_{L}=N_{a}M/L$ is the mass per length. $v_{g}=|v'-v''|$ is the group velocity. The prime over the summation indicates the selection rules imposed on those phonon modes in the summation.

In the low temperature region, the thermal conductivity is dominated by the boundary and defect scattering, as the three-phonon process is very weak. We consider the boundary scattering by the following relaxation rate:\cite{Ziman}
\begin{eqnarray}
\frac{1}{\tau_{\rm bs}} & = & \frac{v_{\kappa,n,z}^{\sigma}}{L}\times\frac{1-p}{1+p},
\label{eq_bslifetime}
\end{eqnarray}
where $p$ is the specularity parameter. For example, in studies of graphene this specularity parameter was fitted to experimental data.\cite{Nika}
\begin{figure}[htpb]
  \begin{center}
    \scalebox{1.0}[1.0]{\includegraphics[width=8cm]{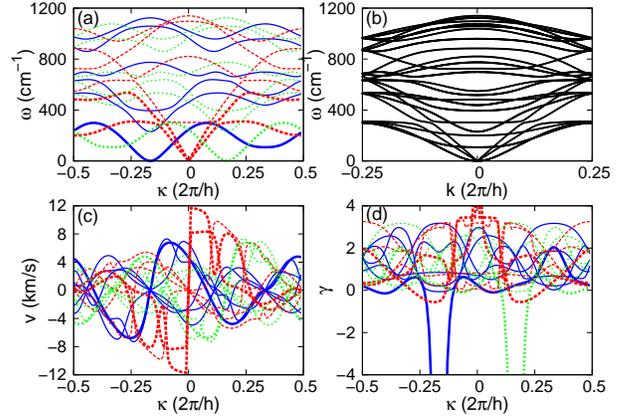}}
  \end{center}
  \caption{(Color online) Physical quantities in SWBNT(3,3). (a). Phonon spectrum parameterized by helical quantum numbers $(\kappa, n)$. (b). Phonon spectrum parameterized by linear quantum numbers $(k,m)$. (c). Phonon velocity. (d). Gr\"uneisen parameter. In (a), (c) and (d), $n=$ -1, 0, 1 are displayed by solid line (blue), dashed line (red), and dot line (green). The four acoustic phonon modes are highlighted by thicker lines.}
  \label{fig_dispersion}
\end{figure}
 Since there is no experiment on thermal boundary scattering phenomenon, we fit this parameter to the molecular dynamics simulation results on short SWBNT of 10 nm long where boundary scattering is important as shown in Fig.~\ref{fig_pfitmd}. The simulation is run based on the same potential. In the molecular dynamics simulation, the thermal current is mimicked by setting two different temperatures on the two ends of the tube. The N\'ose-Hoover\cite{Nose, Hoover} thermostat is employed to maintain constant temperatures with a relaxation time of 0.4 ps. The description of boundary scattering is phenomenological, since we do not have an atomic model for the boundary scattering. Eq.~(\ref{eq_pslifetime}) will give some error in the calculation of $v_{g}$ for short tubes, where the $\kappa$ space is discrete. However, the error will be reasonably small, as the phonon life time in short tubes is mainly determined by the boundary scattering.

 The total phonon life time can be obtained through the Matthiessen's rule:
\begin{eqnarray}
\frac{1}{\tau_{\rm tot}} & = & \frac{1}{\tau_{\rm ps}}+\frac{1}{\tau_{\rm bs}}.
\label{eq_totlifetime}
\end{eqnarray}

From the above, we can calculate the phonon life time and thermal conductivity due to boundary scattering and  three-phonon scattering. This procedure requires the linear properties including phonon spectrum $\omega_{\kappa,n}^{\sigma}$, phonon velocity $v_{\kappa,n}^{\sigma}$, and the nonlinear property of the Gr\"uneisen parameter $\gamma_{\kappa,n}^{\sigma}$. All of these quantities are obtained from the Tersoff+UFFOOP potential.

\begin{figure}[htpb]
  \begin{center}
    \scalebox{1.0}[1.0]{\includegraphics[width=8cm]{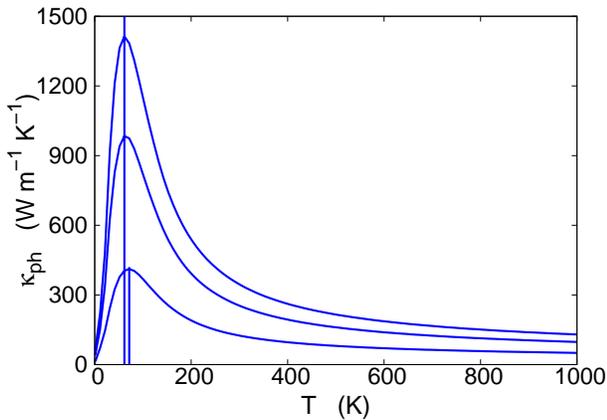}}
  \end{center}
  \caption{(Color online) Thermal conductivity of SWBNT(5,0) with length 1, 5, 10 $\mu$m from bottom to top.}
  \label{fig_cdt_swbnt_5_0_tem}
\end{figure}
\section{calculation results and discussion}
We employ the Tersoff+UFFOOP potential to calculate all physical quantities required in the calculation of thermal conductivity. Fig.~\ref{fig_dispersion} shows the results for SWBNT(3,3), where $\kappa\in(-\pi/h, \pi/h]$ with $h=1.23$~{\AA}. Quantities belonging to $n=$ $-1$, 0, 1 are plotted in solid (blue online), dashed (red online), and dot (green online) lines. Thicker lines are used to mark physical quantities of the four acoustic phonon dispersions, i.e., the LA, the twisting (TW), and two TA phonon modes (or flexure modes). Panel (a) is the phonon spectrum parameterized by the helical quantum numbers $(\kappa,n)$. Different from the LA and TW modes, the zero point of the two TA modes are located at $(\kappa,n)=$ $(-\alpha,-1)$ and $(\alpha,1)$. Panel (b) shows the phonon spectrum parameterized by the linear quantum numbers $(k, m)$. The BZ of $k$ is only half of the BZ for the $\kappa$ and $m=-3,-2,-1,0,1,2$. These curves are obtained through the relationship between two sets of quantum numbers. The zero points of the four acoustic phonon dispersions are all in the $\Gamma$ point $(k,m)=(0,0)$. We can not tell the difference between the TA modes and the other acoustic modes through the position of the zero frequency any more. Panel (c) shows the phonon velocity. Let us concentrate on the six curves attributed to $n=0$ (dashed line, red online). Among the six phonon modes at $\kappa=0$, LA/TW modes are rigid translation/rotation of the SWBNT; so they do not contribute to the thermal conductivity. The other four modes have zero velocity; thus can not carry heat energy during thermal transport. Similarly, the twelve phonon modes at $\kappa=\pm \pi/h$ do not transfer heat energy directly, as their velocities are zero. In other SWBNT with even $N$, eg. SWBNT(4,2), we find that phonon velocities are also zero at BZ boundary $(\kappa,n)=(0,N/2)$.
\begin{figure}[htpb]
  \begin{center}
    \scalebox{0.8}[0.8]{\includegraphics[width=8cm]{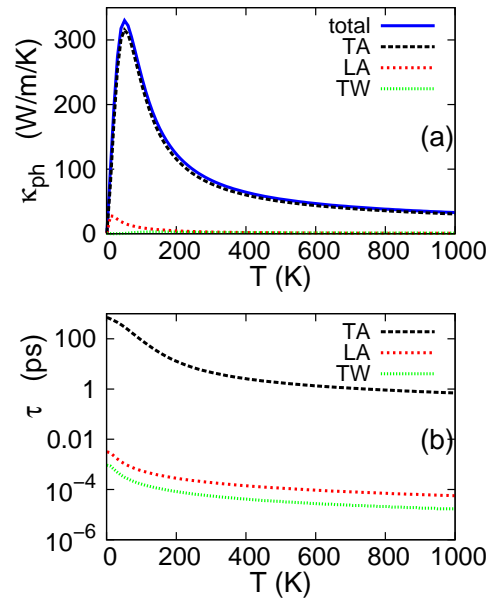}}
  \end{center}
  \caption{(Color online) (a). Thermal conductivity of 1 $\mu$m SWBNT(3,3) contributed from different acoustic phonon modes. (b). Life times for acoustic phonon modes.}
  \label{fig_cdt_swbnt_3_3_mode}
\end{figure}
 Actually, the zero velocity of phonon modes at BZ edge and optical modes at $\Gamma$ point is the result of the continuity and inversion symmetry of phonon velocities for optical modes. This is also true in bulk materials where the selection rules of three-phonon process are merely to conserve the momentum besides the energy conservation, while the conservation of quantum numbers corresponding to other point symmetries is ignored. Panel (d) shows the Gr\"uneisen parameter. Most phonon modes have positive $\gamma$, except the two TA modes whose $\gamma$ are large negative numbers. The large negative Gr\"uneisen parameter was also found in 2D graphene by Mounet {\it et~al.} doing first-principles calculation.\cite{Mounet} It is the origin of thermal contraction at low temperatures in many materials. The obtained Gr\"uneisen parameters are quite different for different phonon modes; so it is important to use a mode-dependent Gr\"uneisen parameter $\gamma_{\kappa,n}^{\sigma}$ in the calculation of thermal conductivity from Eqs.~(\ref{eq_conductivity})~(\ref{eq_pslifetime}).

Figure.~\ref{fig_cdt_swbnt_5_0_tem} shows the thermal conductivity in temperature range [1.0, 1000.0] K for SWBNT(5,0) with several different lengths. In the low temperature region, the thermal conductivity is limited mainly by the boundary scattering. $\kappa_{\rm ph}$ increases with increasing temperature according to the temperature dependence of phonon heat capacity as the boundary scattering is temperature independent. Around $T_{c}=80$ K, thermal conductivity reaches a maximum value, where boundary scattering and three-phonon scattering counterbalance each other. Above $T_{c}$, the three-phonon process becomes more important leading to the decrease of $\kappa_{\rm ph}$ with further temperature increase. The critical temperature $T_{c}$ is lower in longer tubes where boundary scattering contributes less than shorter tubes.
\begin{figure*}[htpb]
  \begin{center}
    \scalebox{1}[1]{\includegraphics[width=\textwidth]{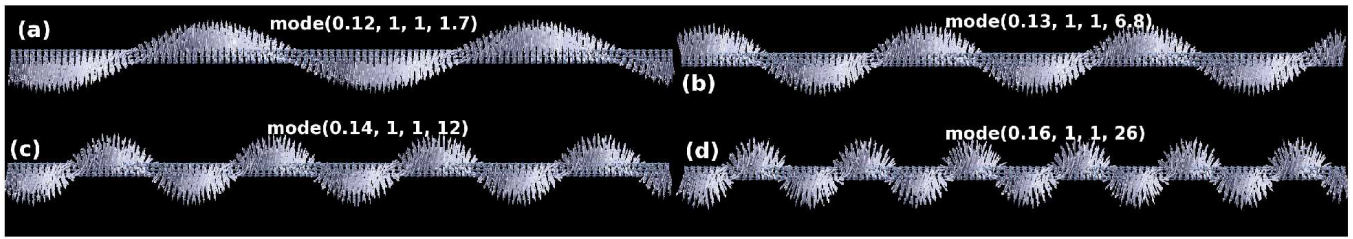}}
  \end{center}
  \caption{The vibrational morphology of four flexure modes in SWBNT(5,0) with frequency increased from 1.7 cm$^{-1}$ to 26 cm$^{-1}$ in panels (a) to (d). The values displayed are $(\kappa,n,\sigma,\omega)$ for each mode.}
  \label{fig_flexuremode}
\end{figure*}
 With the increase of length, $\kappa_{\rm ph}$ increases due to weaker boundary scattering, exhibiting the ballistic characteristic thermal transport.

The acoustic modes are major heat carriers, because of their higher velocities compared with optical modes. Fig.~\ref{fig_cdt_swbnt_3_3_mode}~(a) shows the contribution from the four acoustic modes to $\kappa_{\rm ph}$ in 1 $\mu$m long SWBNT(3,3). It shows that the two TA modes dominant the thermal conductivity, although their velocities are smaller than the LA and TW acoustic modes. The TA vibrations are also most visible in the molecular dynamics simulation. Fig.~\ref{fig_flexuremode} displays the vibrational morphology of four selected TA modes. The length of arrow is proportional to the amplitude of vibrational displacement.

The TA modes are more effectively excited than LA and TW modes in whole temperature range because of their quadratic phonon spectra. As a result, there are more TA phonons to carry the heat energy. However, the TA modes should not have such a dominant contribution to $\kappa_{\rm ph}$ if they can not carry heat energy for a long time. Hence, it is also necessary to compare the life time of different phonon modes. The comparison is shown in Fig.~\ref{fig_cdt_swbnt_3_3_mode}~(b). It shows that the life time of TA mode is about three to four orders longer than the LA/TW modes. From Fig.~\ref{fig_dispersion}~(d), the value of Gr\"uneisen parameters for general TA modes are close to that of the LA/TW modes. It means that there should be almost the same phonon-phonon scattering for all three acoustic modes between phonon modes $(\kappa,n,\sigma)$, $(\kappa',n',\sigma')$, and $(\kappa'',n'',\sigma'')$, satisfying the selection rules. However, for long-wave LA/TW modes, the energy conservation will always lead to $\sigma'=\sigma''$, while the other selection rules require $\kappa'$ to be close to $\kappa''$ and $n'=n''$. From Fig.~\ref{fig_dispersion}~(c), two modes with closer $\kappa$ and the same $(n,\sigma)$ will have similar phonon velocity, which results in very small value of group velocity $v_{g}$ in Eq.~(\ref{eq_pslifetime}). As a result, the three-phonon scattering rate is extremely strong for LA/TW modes. The situation is quite different for TA modes, where the selection rules require $\kappa'$ to be far away from $\kappa''$ and $n\not=n'$. It will lead to a large value of $v_{g}$, which eventually results in a considerably weak three-phonon scattering rate. Considering the above two aspects of TA modes, we learn that the TA modes are fully excited in whole temperature range and can carry heat for a long time; so they dominate the thermal conductivity.

It has been well established that the thermal conductivity in bulk materials will converge with length increasing at some point, yet it is still an open issue in low-dimensional materials such as graphene or nanotube. The thermal conductivity shows different length dependence in different transport region. In the pure ballistic region, the thermal conductivity is proportional to the length; while $\kappa_{\rm ph}$ is length-independent in pure diffusive region. Intuited by these two situations, it is a usual trick to fit the thermal conductivity as an exponential function of length. This technique is useful in the study of thermal conductivity in nano-materials, where the Fourier law fails and the thermal transport is neither ballistic nor diffusive. From this point of view, the beta exponent in length dependence of thermal conductivity is a nice way to observe the thermal transport nature. The origins for the exponential behavior in ballistic and diffusive regions are clear, while there is no good reason for the exponential behavior of thermal transport between ballistic and diffusive. It can be taken as an artifact of fitting approximation at this moment.

We calculate $\kappa_{\rm ph}$ in [1.0, 1000] K for SWBNT(5,0) with lengths distributed in [0.3, 10] $\mu$m and SWBNT(10,0) with length $L\in$[0.3, 5] $\mu$m.
\begin{figure}[htpb]
  \begin{center}
    \scalebox{1.0}[1.0]{\includegraphics[width=8cm]{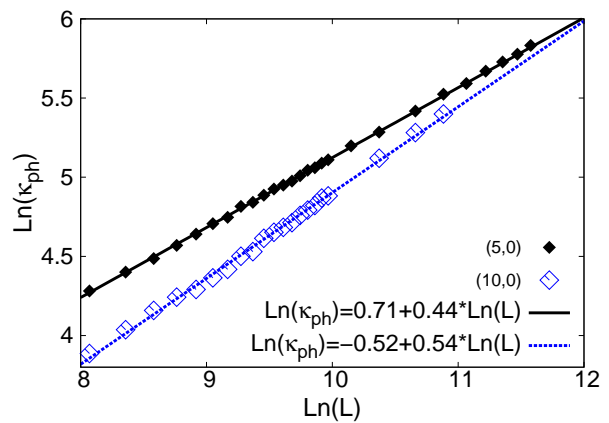}}
  \end{center}
  \caption{(Color online) A natural logarithm of the thermal conductivity and length. Power function fitting for the thermal conductivity of SWBNT(5,0) with $L\in$ [0.3, 10]$\mu$m, and SWBNT(10,0) with $L\in$ [0.3, 5]$\mu$m at 300 K.}
  \label{fig_cdt_swbnt_length}
\end{figure}
\begin{figure}[htpb]
  \begin{center}
    \scalebox{1.0}[1.0]{\includegraphics[width=8cm]{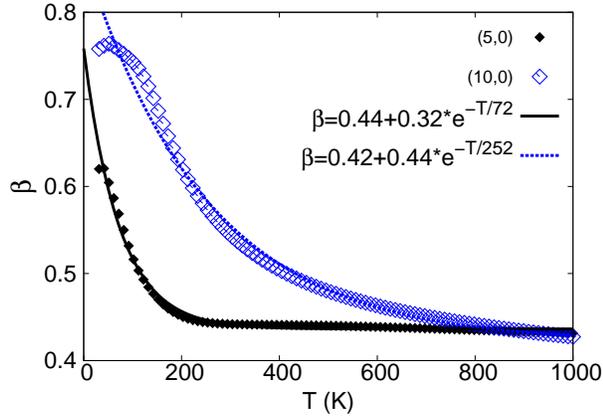}}
  \end{center}
  \caption{(Color online) Temperature dependence for the power factors of SWBNT(5,0) and (10,0).}
  \label{fig_cdt_swbnt_powerfactor}
\end{figure}
 At each temperature, we fit $\kappa_{\rm ph}$ as a power function of $L$: $\kappa_{\rm ph}=aL^{\beta}$. Two fitting results at 300 K are illustrated in Fig.~\ref{fig_cdt_swbnt_length}. In bulk materials, $\beta=1$ for ballistic transport, while $\beta=0$ in case of diffusive transport. For nano materials, the ballistic transport is easier to be observed, because the structure is too small for phonon-phonon scattering to take effect. However, a pure diffusive thermal transport has seldom been seen in quasi-one-dimensional systems.\cite{MaruyamaS,YaoZ,ZhangG,WangJ} $\beta$ should be smaller at higher temperatures where the three-phonon process becomes more important. A detailed study of the temperature-dependence of $\beta$ is very difficult to be done by molecular dynamics simulation because of large computation requirement. The above approach in Sec.IV can be used to perform this calculation efficiently. Fig.~\ref{fig_cdt_swbnt_powerfactor} shows that $\beta$ decreases exponentially with increasing temperature. $\beta$ has large value at low temperatures as it is almost in the ballistic transport region. With temperature increase, $\beta$ decreases faster in thinner tubes. In the high temperature limit, $\beta=$ 0.44 and 0.42 for SWBNT(5,0) and (10,0). The thermal transport in both tubes are not in the pure diffusive region. Thicker tubes are closer to the pure diffusive transport with smaller $\beta$. It is still unclear why the exponent $\beta$ decays exponentially with temperature increasing, even though such behavior has been clearly observed here.

\begin{figure}[htpb]
  \begin{center}
    \scalebox{1}[1]{\includegraphics[width=8cm]{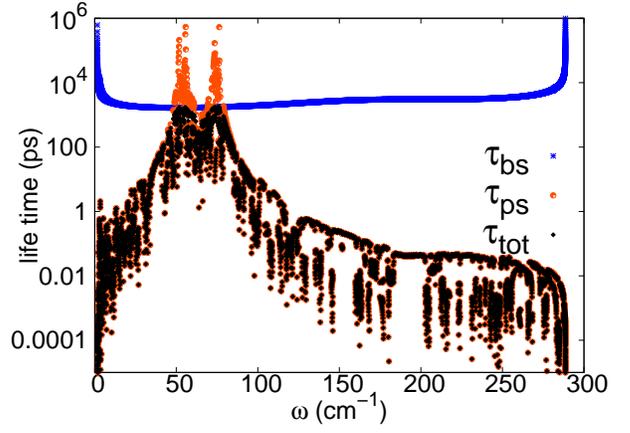}}
  \end{center}
  \caption{(Color online) Phonon life time of TA modes in 2$\mu$m SWBNT(10,0) at 300 K. The life time due to boundary scattering, $\tau_{bs}$, is shown by blue stars. The three-phonon scattering life time, $\tau_{ps}$, is displayed by red open circles. The total life time is shownn by black diamond.}
  \label{fig_lifetime}
\end{figure}
The divergence of thermal conductivity with nonzero $\beta$ is the result of the competition between three-phonon scattering and boundary scattering of the TA modes, which dominate the thermal conductivity. Fig.~\ref{fig_lifetime} shows the life time for TA modes at room temperature. The drop-off of life time for low frequency phonons is due to the quadratic dispersion of the flexure modes, which leads to extremely large Gr\"uneisen parameter of these modes around $\kappa=\pm\alpha$ as displayed by Fig.~\ref{fig_dispersion}~(d). Another important feature of Fig.~\ref{fig_lifetime} is that the three-phonon scattering for TA modes around 50 and 70 cm$^{-1}$ is so weak that the life times of these modes are mainly limited by the boundary scattering. This part of contribution to the thermal conductivity will diverge with increasing length, i.e $\beta$ should be 1.0. Other flexure modes have considerable three-phonon scattering. Contribution from these phonons to the thermal conductivity does not depend on the length of SWBNT, i.e $\beta$ should be 0. The counteraction between these two mechanisms results in a power factor $\beta$ in the range [$0$, $1$]. Actually, a similar phenomenon was also found by Nika {\it et.al} for the thermal conductivity in graphene, where the boundary scattering still plays an important role for very large piece of graphene sample and eventually leads to the increase of thermal conductivity with increasing size.\cite{Nika}

\section{conclusion}
To conclude, the present work calculates the phonon life time due to boundary scattering and three-phonon scattering process in SWBNT. The linear and nonlinear physical quantities required in the calculation are obtained from the Tersoff+UFFOOP inter-atomic potential, which inherits the efficiency and stability of the original Tersoff potential and is suitable for the field of heat transport. The selection rules for three-phonon process are figured out by analyzing the symmetry group (line group) of SWBNT. A set of helical quantum numbers $(\kappa,n)$ corresponding to the line group is accepted in the selection rules instead of the usual linear quantum numbers $(k,m)$ corresponding to a subgroup of the line group. The calculation is focusing on the thermal conductivity for SWBNT with different lengths and the contribution of different phonon modes in the heat transport.

\textbf{Acknowledgements} The authors thank Y. F. Gu and Prof. Y. F. Chen at Southeast University for helpful correspondence, and Prof. B. S. Wang at IOS-CAS for insightful discussions. The work is supported by a URC grant of R-144-000-257-112 of National University of Singapore.

\end{document}